\documentclass{aa}
\usepackage{graphicx}
\usepackage{multirow}
\usepackage{aalongtable}

\begin{document}

\title{Multiwavelength monitoring of \object{BD~$+53\degr$2790}, the optical counterpart to \object{4U~2206+54}}

\author{ P.~Blay\inst{1}
\and  I.~Negueruela\inst{2}             
\and P.~Reig\inst{3,4}
\and M.~J.~Coe\inst{5}
\and R.~H.~D.~Corbet\inst{6,7}       
\and J.~Fabregat\inst{8}
\and A.~E.~Tarasov\inst{9}}

\offprints{P. Blay, \email{pere.blay@uv.es}}

\institute{Institut de Ci\`encia dels Materials, Universidad de Valencia, P.O. BOX 22085, E46071 Valencia, Spain\\
\email{pere.blay@uv.es}
\and Departamento de F\'{\i}sica, Ingenier\'{\i}a de Sistemas y Teor\'{\i}a de la Se\~{n}al, EPSA, Universidad de Alicante, P.O. BOX 99, E03080, Alicante, Spain\\
\email{ignacio@dfists.ua.es}
\and Foundation for Research and Technology-Hellas, 711 10, Heraklion, Crete, Greece
\and Physics Department, University of Crete, 710 03 Heraklion, Crete, Greece\\
\email{pau@physics.uoc.gr}
\and School of Physics and Astronomy, Southampton University, Southampton, SO17 1BJ, U.K.\\
\email{M.J.Coe@soton.ac.uk}
\and X-ray Astrophysics Laboratory, Code 662, NASA/Goddard Space Flight Center, Greenbelt, MD 20771, U.S.A.
\and Universities Space Research Association\\
\email{corbet@milkyway.gsfc.nasa.gov}
\and Observatori Astron\'omic, Universidad de Valencia, P.O. BOX 22085, E46071 Valencia, Spain\\
\email{juan@pleione.oauv.uv.es}
\and Crimean Astrophysical Observatory, Nauchny, Crimea, 334413, Ukraine\\
\email{tarasov@crao.crimea.ua}}

\date{Received / Accepted}

\abstract{We present the results of our long-term monitoring of 
\object{BD~$+53\degr$2790}, the optical counterpart to the X-ray source \object{4U~2206+54.} 
Unlike previous studies that classify the source as a Be/X-ray binary, we find that its
optical and infrared properties differ from those of typical Be stars: the variability of the V/R 
ratio is not cyclical; there are variations in the shape and strength of the H$\alpha$ emission line on timescales 
less than 1 day; and no correlation between the EW and the IR magnitudes or colors is seen. Our observations
suggest that \object{BD~$+53\degr$2790} is very likely a peculiar O9.5V star.
In spite of exhaustive searches we cannot find any significant modulation in any emission line parameter
or optical/infrared magnitudes. Spectroscopy of the source extending from the optical to the 
$K$-band confirms the peculiarity of the spectrum: not only are the He lines stronger than expected for 
an O9.5V star but also there is no clear pattern of variability. The possibility that
 \object{BD~$+53\degr$2790}
is an early-type analogue to He-strong stars (like $\theta^1$ Ori C) is discussed.
}

\maketitle

\section{Introduction}

4U\,2206+54, first detected by 
the {\em UHURU} satellite (Giacconi et al. 1972), is a weak persistent
X-ray source. It has been observed by {\em Ariel V} (as 3A\,2206+543; 
Warwick et al. 1981), {\em HEAO--1}
(Steiner et al. 1984), {\em EXOSAT} (Saraswat \& Apparao 
1992), {\em ROSAT} (as 1RX\, J220755+543111; Voges et al. 1999), {\em RossiXTE}
(Corbet \& Peele, 2001; Negueruela \& Reig 2001, henceforth NR01) and {\em INTEGRAL}
(Blay et al., 2005 ). The source is variable, by a factor $>3$ on timescales
of a few minutes and by a factor $>10$ on longer timescales (Saraswat 
\& Apparao 1992; Blay et al. 2005), keeping an average luminosity around 
$L_{{\rm x}} \approx 10^{35}\:{\rm erg}\,{\rm s}^{-1}$
for an assumed distance of $3\:{\rm kpc}$ (NR01).

The optical counterpart was identified by Steiner et al. (1984), based
on the position from the {\em HEAO--1} Scanning Modulation Collimator, as the early-type
star \object{BD~$+53\degr$2790}. The star displayed H$\alpha$ line in 
emission with two clearly differentiated peaks, separated by about 460 km s$^{-1}$. Even though some characteristics
of the counterpart suggested a Be star (Steiner et al., 1984), high resolution
spectra show it to be an unusually active O-type star, with an
approximate spectral type O9Vp (NR01). 

{\em RossiXTE}/ASM observations of \object{4U~2206+54}, show the
X-ray flux to be modulated with a period of approximately 9.6 days (see 
Corbet \& Peele, 2001; Rib\'o et al., 2005). The short orbital period, absence of X-ray pulsations
and peculiar optical counterpart make  \object{4U~2206+54} a rather
unusual High-Mass X-ray Binary (HMXB). The absence of pulsations indicates that the compact
companion could be a black hole. Recent studies of high energy emission from the system, however,
suggest that the compact object in \object{4U42206+54} is a neutron star
(Blay et al., 2005; Torrej\'on et al. 2004; Masseti et al. 2004).

In an attempt to improve our knowledge of this system, we have collected optical and infrared
observations covering about 14 years. 

\section{Observations}

We present data obtained as a part of a long-term monitoring campaign
consisting of optical and infrared spectra, infrared and optical 
broad-band photometry and narrow-band Str\"{o}mgren optical photometry of
\object{BD~$+53\degr$2790}, the optical counterpart to \object{4U~2206+54}.

\subsection{Spectroscopy}

\subsubsection{Optical Spectroscopy}

We have monitored the source from 1990 to 1998, using the 2.5-m
Isaac Newton Telescope (INT) and the 1.0-m Jakobus Kapteyn Telescope
(JKT), both located at the Observatorio del Roque de los
Muchachos, La Palma, Spain, and the 1.5-m telescope
at Palomar Mountain (PAL). We have also made use of data from
the La Palma Archive (Zuiderwijk et al. 1994). The archival data
consist of H$\alpha$ spectroscopic observations taken with the INT
over the period 1986\,--\,1990. The two datasets overlap for a few months 
and together they constitute continuous coverage of the source
for thirteen years. The older INT observations had been taken with
the Intermediate Dispersion Spectrograph (IDS) and 
either the Image Photon Counting System (IPCS) or a CCD
camera. All the INT data after 1991 were obtained with CCD cameras.
The JKT observations were obtained using the St Andrew's
Richardson-Brealey Spectrograph
(RBS) with the R1200Y grating, the red optics and either the EEV7 or
TEK4 CCD cameras, giving a nominal dispersion of $\approx$ 1.2 \AA. The
Palomar 1.5-m was operated using the f/8.75 Cassegrain echelle
spectrograph in regular grating mode (dispersion $\approx0.8$\AA/pixel).

Further observations were taken with the 2.6-m telescope at the
Crimean Astrophysical Observatory (CRAO) in Ukraine.

From 1999, further monitoring has been carried out using the 1.52-m
G.~D.~Cassini telescope at the Loiano Observatory (BOL),
Italy, equipped with the Bologne Faint Object Spectrograph and Camera
(BFOSC) and the 1.3-m Telescope at the Skinakas Observatory (SKI), in 
Crete, Greece. From Loiano, several observations were taken using 
grism\#8, while higher resolution spectra were taken with grism\#9 in
echelle mode (using grism\#10 as cross-disperser). Other spectra were
taken with the echelle mode of grism\#9 and grism\#13 as
cross-disperser, giving coverage of the red/far-red/near-IR region (up
to $\sim 9000\,$\AA). At Skinakas, the  telescope is an
f/7.7 Ritchey-Cretien, which was equipped with a $2000\times800$ ISA SITe
chip CCD and a 1201~line~mm$^{-1}$ grating, giving a nominal dispersion of
1~\AA~pixel$^{-1}$.

Blue-end spectra of the source have also been taken with all the
telescopes listed, generally using the 
same configurations as in the red spectroscopy, but with blue gratings 
and/or optics when the difference was relevant (for example, from
Loiano, grisms \#6 and \#7 were used for the blue and yellow regions 
respectively).

All the data have been reduced using the {\em Starlink}
software package {\sc figaro} (Shortridge et al., \cite{shortridge}) and
analysed using {\sc dipso} (Howarth et al., \cite{howarth97}). Table \ref{tab:log}
lists a log of the spectroscopic observations.

\subsubsection{Infrared Spectroscopy}

Near-infrared ($I$ band) spectra of \object{BD~$+53\degr$2790} have also
been taken with the JKT, INT and G.~D.~Cassini telescopes.

{\em K}-band spectroscopy of \object{BD~$+53\degr$2790} was obtained
on July 7-8, 1994, with the Cooled Grating Spectrometer (CGS4) on UKIRT,
Hawaii. The instrumental configuration consisted of the long focal
station (300 mm) camera and the 75 lines\,mm$^{-1}$ grating, which gives a
nominal velocity resolution of 445 km\,s$^{-1}$ at 2$\mu$m
($\lambda/\Delta \lambda \approx 700$). The data were reduced according
to the procedure outlined by \cite{eve93}.

\subsection{Photometry}
\subsubsection{Optical Photometry}

We took one set of {\em UBVRI} photometry of the source on August 18,
1994, using the 1.0-m Jakobus Kapteyn Telescope (JKT). The observations 
were made using the TEK\#4
CCD Camera and the Harris filter set. The data have been calibrated
with observations of photometric standards from \cite{landolt92} and the
resulting magnitudes are on the Cousins system. 

We also obtained several sets of Str\"{o}mgren {\em uvby}$\beta$
photometry. The early observations were taken at the 1.5-m 
Spanish telescope at the German-Spanish Calar Alto Observatory, Almer\'{\i}a, 
Spain, using the {\em UBVRI}\, photometer with the $uvby$ filters, in 
single-channel mode, attached to the Cassegrain focus. Three other sets were
obtained with the 1.23-m telescope at Calar Alto, using the TEK\#6 CCD
equipment. One further set was taken with the 1.5-m 
Spanish telescope equipped with the single-channel multipurpose photoelectric 
photometer. Finally, one set was obtained with the 1.3-m Telescope at
Skinakas, equipped with a Tektronik $1024\times1024$ CCD.

%----------------------------------------------------------------------
\begin{table*}[h!]
\caption{Str\"{o}mgren photometry of the optical counterpart to
4U\,2206+54. The last column 
indicates the telescope used. a stands for the 1.5-m Spanish telescope
at Calar Alto. b represents the 1.23-m German telescope. c is the
Skinakas 1.3-m telescope.}
\label{tab:opticalphotom}
\begin{center}
\begin{tabular}{lccccccc}
\hline\hline
Date & MJD &$V$ &$(b-y)$ & $m_{1}$ &$c_{1}$ &$\beta$&T\\
\hline
 & & & & & & & \\
1988, Jan 7 &47168.290   & 9.909$\pm$0.013  &0.257$\pm$0.005  &$-$0.083$\pm$0.007  & 0.011$\pm$0.007    &2.543$\pm$0.040 & a \\
1989, Jan 4 &74531.305   & 9.845$\pm$0.015  &0.257$\pm$0.007  &$-$0.042$\pm$0.010  & $-$0.117$\pm$0.017 &2.543$\pm$0.007 & a \\
1991, Nov 16 &48577.401  & 9.960$\pm$0.034  &0.268$\pm$0.005  &$-$0.040$\pm$0.012  & $-$0.041$\pm$0.033 & ---            & b \\
1991, Dec 19 &48610.297  & 9.969$\pm$0.038  &0.271$\pm$0.021  &$-$0.322$\pm$0.006  & $-$0.010$\pm$0.018 &2.489$\pm$0.024 & b \\
1994, Jun 21 &49524.500  & 9.835$\pm$0.019  &0.258$\pm$0.013  &$-$0.032$\pm$0.021  & 0.053$\pm$0.030    &2.617$\pm$0.020 & b \\ 
1996, May 26 &50229.642  & 9.845$\pm$0.012  &0.267$\pm$0.007  &$-$0.052$\pm$0.012  & $-$0.074$\pm$0.013 &2.553$\pm$0.006 & a \\
1999, Aug 16 & 51407.500 & 9.883$\pm$0.031  &0.255$\pm$0.044  &$-$0.226$\pm$0.074  & 0.298$\pm$0.094    & $-$            & c \\
\hline
\end{tabular}
\end{center}
\end{table*}
%----------------------------------------------------------------------

All observations are listed in Table \ref{tab:opticalphotom}.

\subsubsection{Infrared Photometry}

Infrared observations of \object{BD~$+53\degr$2790} have been obtained with  
the Continuously Variable Filter (CVF) on the 1.5-m. Carlos S\'{a}nchez
Telescope (TCS) at the Teide Observatory, Tenerife, Spain and the UKT9
detector at the 3.9-m UK Infrared Telescope (UKIRT) on Hawaii. 
All the observations are listed in
Table~\ref{tab:observations}. The errors are much smaller after 1993,
when we started implementing the multi-campaign reduction procedure 
described by Manfroid (\cite{manfroid93}).

\section{Long-term monitoring}

\subsection{Spectrum description and variability}
\begin{figure*}
\begin{centering}
\resizebox{0.9\hsize}{!}{\includegraphics[angle=-90]{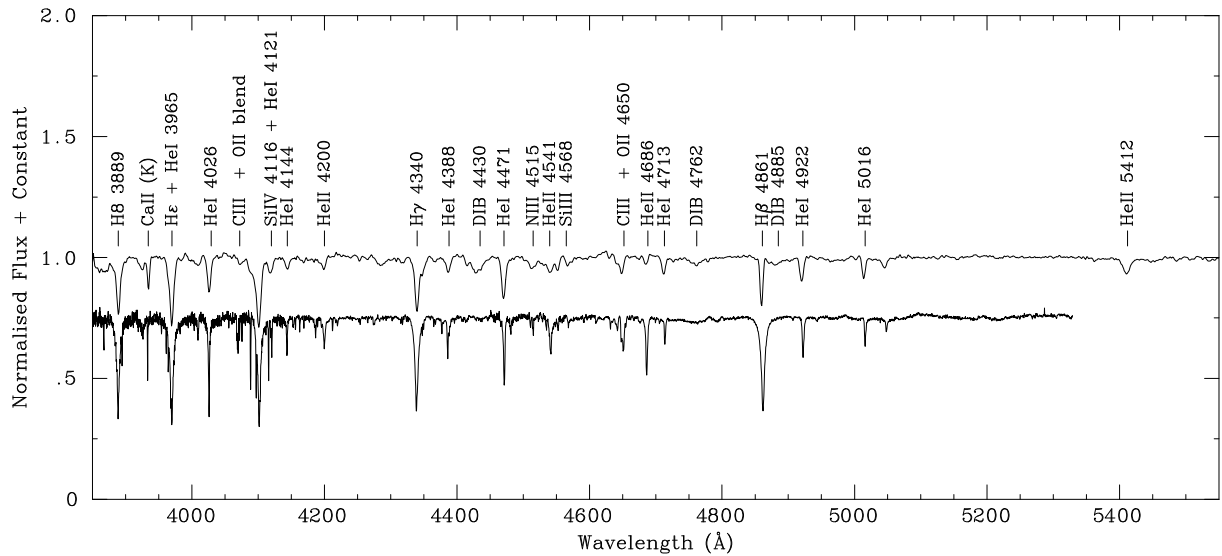}}
\caption{Blue/green spectrum of \object{BD~$+53\degr$2790}, taken on July
    21, 2000 with the 1.3-m telescope at Skinakas. Only the
    strongest features have been indicated. For a more complete
    listing of photospheric features visible in the spectrum, see
    NR01. The spectrum has been normalised by dividing a
    spline fit into the continuum. A normalised spectrum of 10 Lac (09V), shifted down
    for plotting purposes, is also shown for comparison.}
\label{fig:bluegreen}
\end{centering} 
\end{figure*}

\ref{baddata}
Spectra in the classification region (4000--5000~\AA) show all Balmer and He\,{\sc i}
lines in absorption. Several spectra of \object{BD$+53\degr$2790} at 
moderately high resolution were presented 
in NR01, together with a
detailed discussion of its spectral peculiarities. A representative
spectrum covering a wider spectral range is given in
Fig.~\ref{fig:bluegreen}. The rather strong
\ion{He}{ii}~$\lambda$5412\AA\ line represents further confirmation
that the underlying spectrum is that of an O-type star. Together
with  the blue spectrum of \object{BD$+53\degr$2790} a spectrum of
the O9V standard \object{10 Lac} is also shown in Fig.~\ref{fig:bluegreen}.

%---------------------------------------------------------------------------------
%---------------------------------------------------------------------------------

There is no evidence for variability in what can be considered with
certainty to be photospheric features (i.e., the Balmer lines from
H$\gamma$ and higher and all \ion{He}{i} and \ion{He}{ii} lines in the
blue). However, it must be noted that the EW of
H$\gamma$ is $\approx2.2$~\AA\ in all our spectra (and this value should 
also include the blended O\,{\sc ii} $\lambda$4350 \AA\ line), which is too low for any
main sequence or giant star in the OB spectral range (Balona \&
Crampton 1974). Average values of EWs for different lines are indicated in
Table~\ref{tab:ews}. The main spectral type discriminant for O-type stars is the ratio
\ion{He}{ii}~4541\AA/\ion{He}{i}~4471\AA. The quantitative criteria
of \cite{conti71}, revised by \cite{mathys88}, indicate that
\object{BD~$+33\degr$2790} is an O9.5\,V star, close to the limit with O9\,V.

%------------------------------------------------------------

\begin{table}
\caption{Measurement of the EW of strong absorption lines (without 
obvious variability and presumably photospheric) in the spectrum of 
\object{BD~$+53\degr$2790}.}
\label{tab:ews}
\begin{center}
\begin{tabular}{lc}
\hline\hline
Line & EW (\AA)\\
\hline
 & \\
\ion{He}{ii}~$\lambda$4200\AA & 0.4\\
H$\gamma$ & 2.2\\
\ion{He}{i}~$\lambda$4471\AA & 1.3\\
\ion{He}{ii}~$\lambda$4541\AA & 0.4\\
\ion{He}{i}~$\lambda$4713\AA & 0.5\\
\ion{He}{i}~$\lambda$4923\AA & 0.7\\
\end{tabular}
\end{center}
\end{table}
%------------------------------------------------------------

%------------------------------------------------------------

\begin{figure*} 
\begin{centering}
\resizebox{0.9\hsize}{!}{\includegraphics[width=17cm,angle=-90]{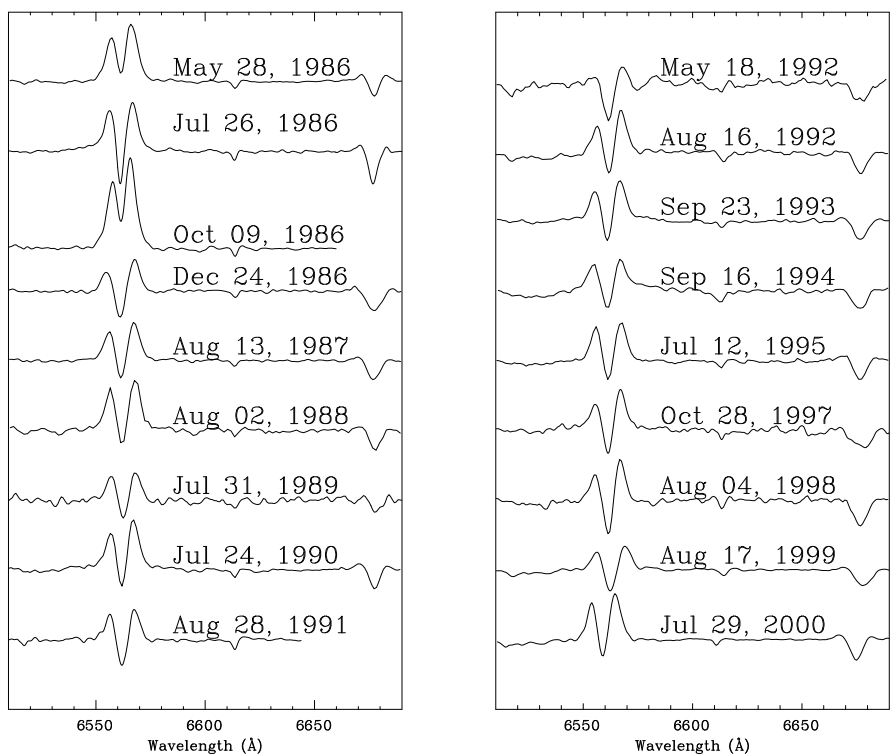}}
\caption{Evolution of the H$\alpha$ line profile of
BD\,$+53^{\circ}$2790 during 1986\,--\,2000. All
spectra have had the continuum level normalised and are offset 
vertically to allow direct comparison.}
\label{fig:halpha}
\end{centering}
\end{figure*}
%------------------------------------------------------------

 Representative shapes
of the H$\alpha$ line in \object{BD~$+53\degr$2790} are shown in
Fig.~\ref{fig:halpha}. 
In all the spectra, two emission components appear clearly
separated by a deep narrow central reversal. The absorption component
normally extends well below the local continuum level -- which is
usually referred to as a ``shell'' spectrum -- but in some spectra,
it does not reach the continuum. The red (R) peak is always stronger
than the blue (V) peak, but the V/R ratio is variable.

The first case of observed strong
variability happened during 1986, when the profile was observed to have
changed repeatedly over a few months from a shell structure to a
double-peaked feature, with the central absorption not reaching the
continuum level. The second one took place in 1992, when the strength
of the emission peaks decreased considerably to about the continuum
level. Finally, during the summer of 2000, we again saw line profiles
in which the central absorption hardly reached the continuum level
alternating with more  pronounced  shell-like profiles.

\begin{figure}[b!]
\centering
\resizebox{0.7\hsize}{!}{\includegraphics[angle=-90]{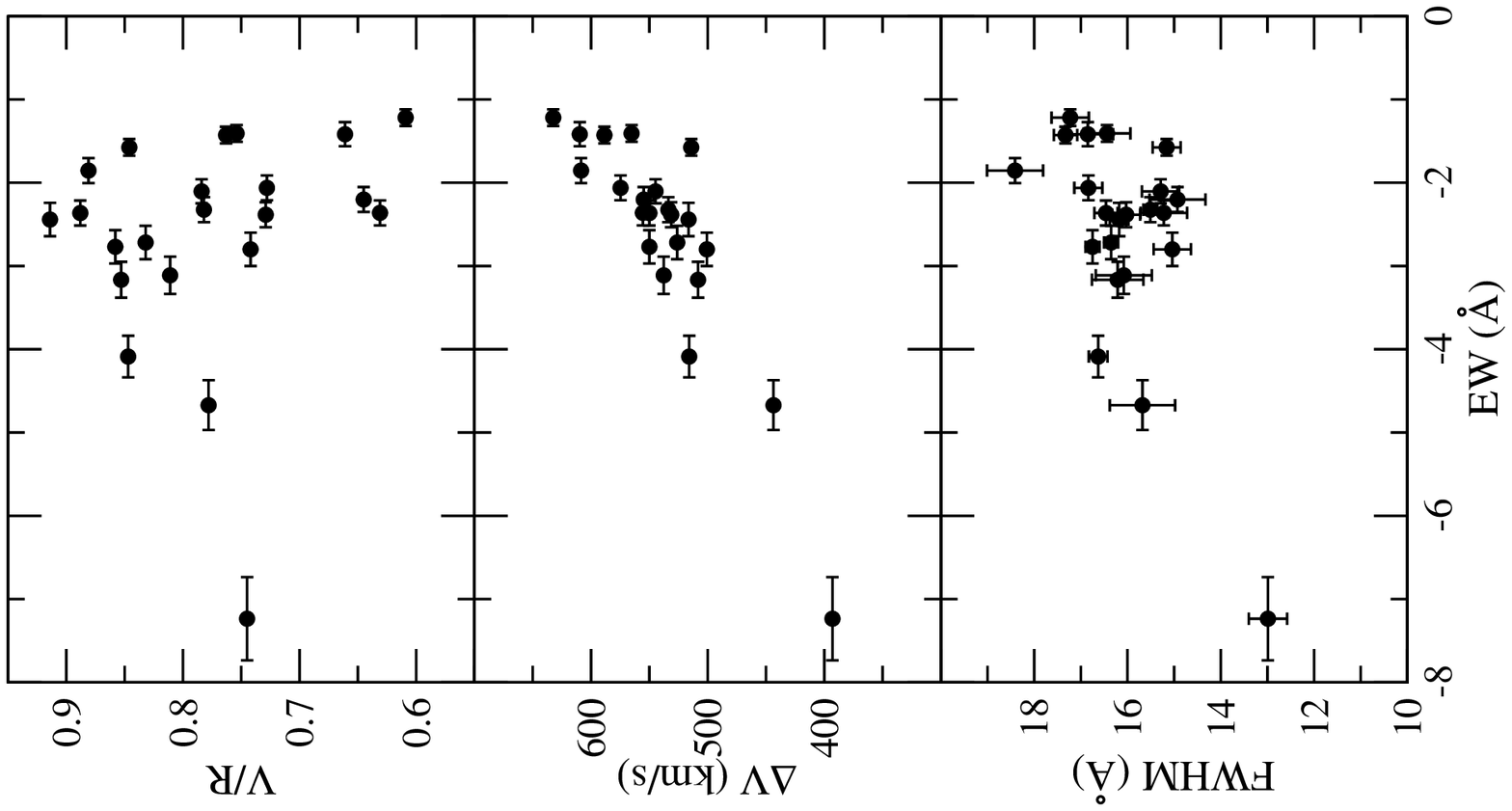}}
\caption{Parameters of the H$\alpha$ emission line for all the red spectra from the INT. }
\label{fig:linepar}
\end{figure}

Figure~\ref{fig:linepar} displays a plot of the Full Width at Half Maximum (FWHM), 
V/R and peak separation ($\Delta$V) of the H$\alpha$ line against its  EW, for all the data from the INT. 
H$\alpha$ parameters (EW, FWHM, V/R and $\Delta$V) were obtained for all the datasets 
shown in Table \ref{tab:log}. Given the very diverse origins of the spectra 
and their very different spectral resolutions, it is difficult to compare them all, 
as there are some effects which introduce some artificial scattering in the data. This 
is the case of the instrumental broadening affecting the FWHM. 
At a first  approximation we considered that it was not necessary to account for it.
Taking into  account the typical spectral resolutions of our dataset --better than 3~\AA~in 
most cases-- and the fact that for the majority of our spectra FWHM $>$ 11~\AA (and generally 
$\approx 14$\AA), the  instrumental broadening, a priori, can be considered negligible. 
\cite{dachs86} found a correlation between  H$\alpha$ parameters (FWHM, peak separation,
EW) in Be stars. We fail to see these correlations when the entire set of
spectra is used but they are present when we restrict the analysis to those
spectra taken with the same instrument, see Fig. \ref{fig:linepar}. There 
is, however, a large spread in the case of the V/R ratio. Most of the scatter in FWHM may be related 
to the larger uncertainties involved when the emission components are small and the line profile is separated.

%However, this might not be the case as in a rotationally dominated disc we
%would expect to find correlations (Dachs et al. 1986, A\&A, 159, 276),
%between the spectral parameters of the H$\alpha$ line (width, peak separation,
%equivalent width). We fail to see those correlations when the entire set of
%spectra is used but are present when we restrict the analysis to those
%spectra taken with the same instrument configuration, see Fig. \ref{fig:linepar}. We find correlation 
%also between the H$\alpha$ EW and the peak separation $\Delta$V, but there is a big 
%spread in the case of the V/R ratio. Most of the scatter in FWHM may be related to the larger 
%uncertainties involved when the emission components are small and the profile split.

\begin{figure}[b!]
\begin{centering}
\resizebox{\hsize}{!}{\includegraphics[angle=-90]{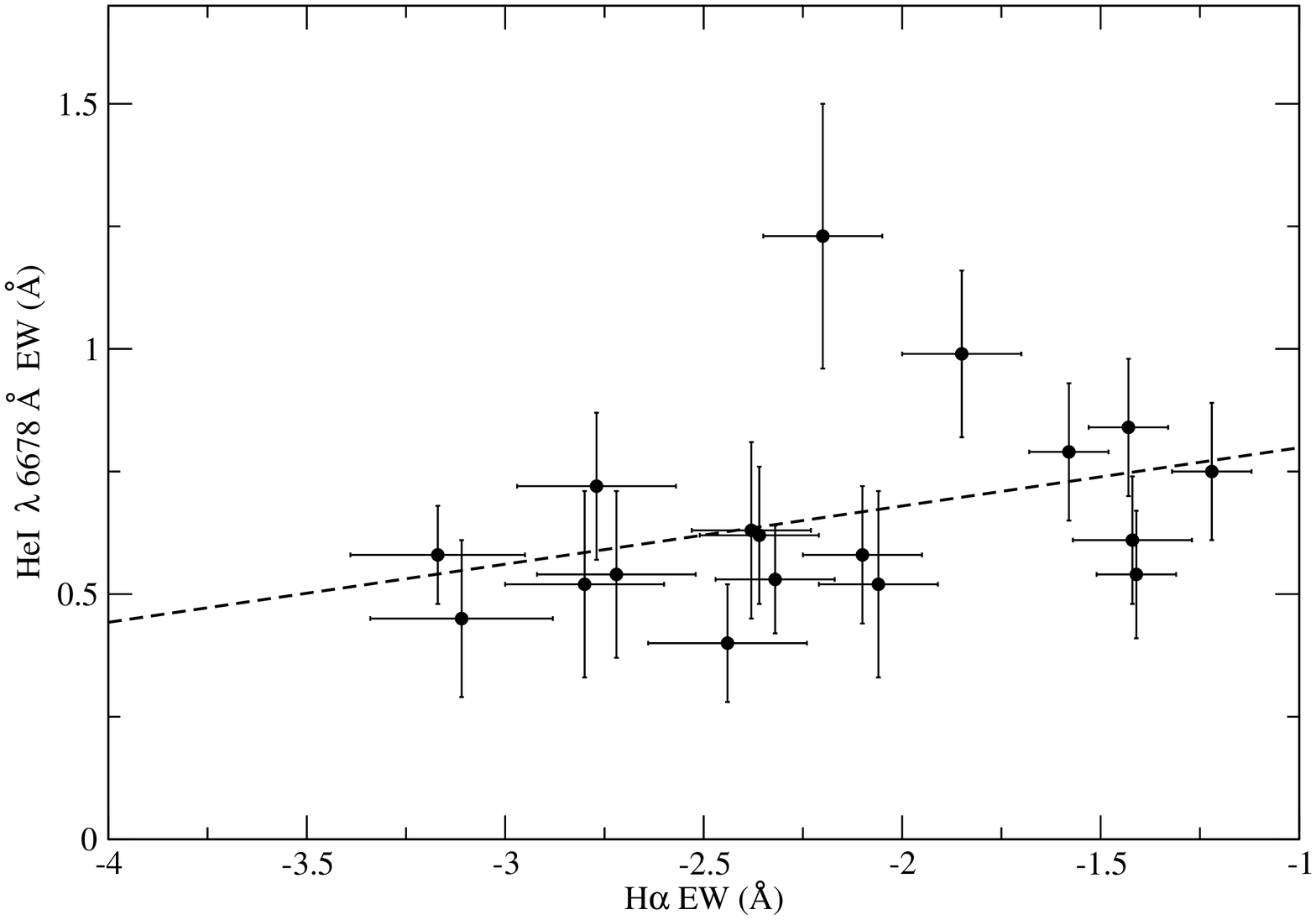}}
\caption{EW of the He\,{\sc i}~$\lambda$6678\AA\ line versus that of the H$\alpha$
line. There seems to be some degree of correlation between both quantities. Only data
from INT spectra where both lines were visible are shown. A linear regresion fit to the data is shown
as a dashed line. The correlation coefficient of the regression is r=0.62 and the correlation
is significant at a 98\% confidence level.} 
\label{fig:halpha_vs_hei} 
\end{centering}
\end{figure}

\begin{figure*} 
\begin{centering}
\resizebox{0.6\hsize}{!}{\includegraphics{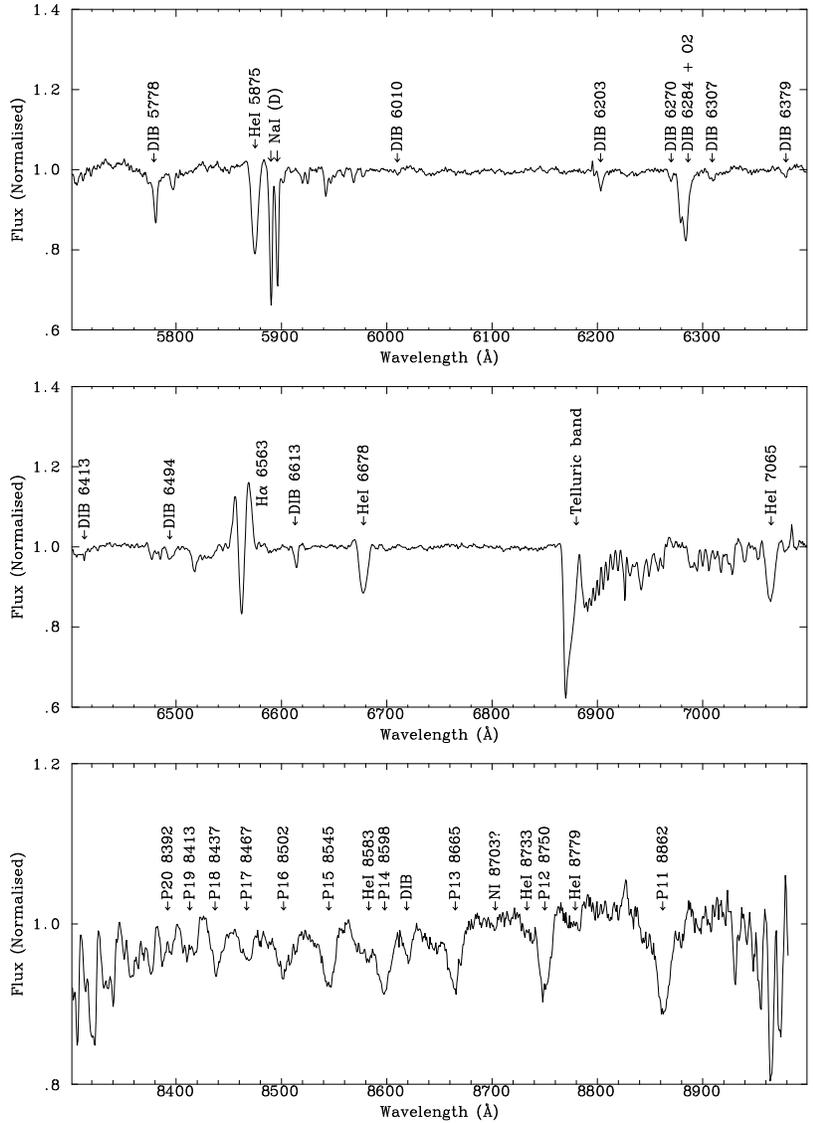}}
\caption{The spectrum of BD~+$53^{\circ}$2790 in the
yellow/red/near-IR. Echelle spectrum taken on 17th August 1999 using
the 1.52-m G.~D.~Cassini Telescope equipped with BFOSC and grisms \#9
(echelle) and \#13 (cross-disperser). All the orders have been
flattened by division into a spline fit to the continuum.} 
\label{fig:spectrum} 
\end{centering}
\end{figure*}

Red spectra covering a larger wavelength range (such as that in 
Fig.~\ref{fig:spectrum}) show also the He\,{\sc i}~$\lambda$\,6678~\AA\ line
and sometimes the He\,{\sc i}~$\lambda$~7065~\AA\ line. Like H$\alpha$, 
the  He\,{\sc i}~$\lambda$\,6678~\AA\  line typically 
displays a shell profile, but the emission peaks are weaker than those
of H$\alpha$, while the central absorption component is normally very
deep. Variability in this line is also more frequent than in
H$\alpha$. The V peak is generally dominant, but the two peaks can be
of approximately equal intensities and sometimes so weak that they
cannot be distinguished from the continuum. Given the apparent different
behaviour of H$\alpha$ and He\,{\sc i}~$\lambda$~6678\AA\ lines, it is 
surprising to find that there is some degree of correlation between their
parameters, as can be seen in Fig. \ref{fig:halpha_vs_hei}, where EW of both
lines from INT spectra in which both lines were visible are shown. 

%--------------------------------------------------------------------
%------------------------------------------------------------------

%---------------------------------------------------------------------------------
%---------------------------------------------------------------------------------

The upper Paschen series lines are always seen in
absorption and no variability is obvious (see Fig. \ref{fig:spectrum}). 
The Paschen lines are much deeper  and narrower than those observed 
in main-sequence OB stars by \cite{and95} and rather resemble
early B-type supergiant stars. 
However, it must be noted that some shell stars in the low-resolution catalogue of
\cite{and88} display $I$-band  spectra that share some characteristics with 
that of \object{BD~$+53\degr$2790}.

\begin{figure} 
\begin{centering}
\resizebox{0.9\hsize}{!}{\includegraphics[]{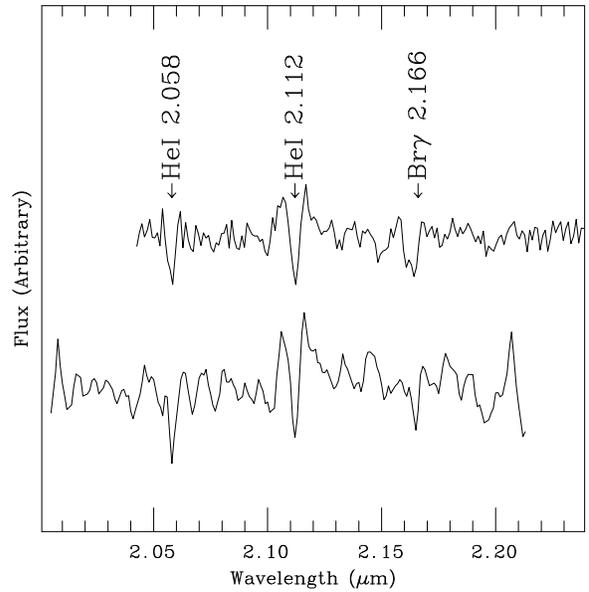}}
\caption{$K$-band spectra of \object{BD~$+53\degr$2790}. The top spectrum was taken on
July 7 1994, and the bottom one on July 8 1994.} 
\label{fig:infra} 
\end{centering}
\end{figure}

$K$-band spectra are shown in Fig. \ref{fig:infra}. Unlike the OB components of several Be/X-ray
binaries observed  by Everall et al. (\cite{eve93}; see also Everall  \cite{eve95}), \object{BD~$+53\degr$2790} 
shows no emission in He\,{\sc i}~$\lambda$2.058 $\mu$m (though the
higher resolution spectrum suggests a weak shell profile). Br$\gamma$
may have some emission component, but is certainly not in emission.
The situation differs considerably from that seen in the $K$-band
spectrum of \object{BD~$+53\degr$2790}  presented by Clark et
al. (1999), taken on 1996 October. There Br$\gamma$ displays a clear
shell profile with two emission peaks and He\,{\sc i} $\lambda$2.112
$\mu$m is in absorption. This shows that the shell-like behaviour and
variability extends into the IR.

%---------------------------------------------------------------------------------
%---------------------------------------------------------------------------------

%------------------------------------------------------------
%------------------------------------------------------------

\subsection{Photometric evolution and colours}

The {\em UBVRI} photometric values we obtain
are $U=9.49$, $B=10.16$, $V=9.89$, $R=9.88$ and $I=9.55$.
The photometric errors are typically 0.05 mag, derived from the
estimated uncertainties in the zero-point calibration and colour
correction. Table \ref{tab:opticalphotom} lists our 
Str\"{o}mgren {\em uvby}$\beta$ measurements.

$V$ measurements in the literature are scarce and consistent with being constant (see
references in NR01). However, our more  accurate set of measurements of the $V$ magnitude 
(or Str\"{o}mgren $y$) show  variability, with a
difference between the most extreme values of $0.13\pm0.05$ mag
(see Table \ref{tab:opticalphotom}), $0.05$ mag being also the standard deviation of 
all 7 measurements.

%Measurements of the $V$ magnitude (or Str\"{o}mgren $y$) for 
%\object{BD~$+53\degr$2790} show little variability. $V$ measurements
%in the literature are consistent with no changes (see
%references in NR01). Our more 
%accurate $uvby$ photometry shows very little sign of variability, with a
%difference between the most extreme values of only $0.13\pm0.05$ mag
%($0.05$ mag is also the standard deviation of all 7 measurements). 

From our $UBV$ photometry, we find that the reddening-free
parameter  $Q$ ($Q=-0.72(B-V)+(U-B)$)is $Q = -0.86\pm0.10$. This value corresponds, according to the
revised $Q$ values for Be and shell stars calculated by Halbedel
(1993), to a B1 star.

We have tried deriving the intrinsic parameters of \object{BD
$+53\degr$2790} from our Str\"{o}mgren photometry by applying the
iterative procedure of \cite{sho83} for de-reddening. The values
obtained for the reddening from the different measurements agree quite
well to $E(b-y)=0.38\pm0.02$ (one standard deviation) and  the
colour $(b-y)_{0}$ averages to $-0.12\pm0.02$. This value corresponds to a B1V star
according to the calibrations of \cite{per87} and \cite{pop80}.

\begin{figure} 
\centering
\resizebox{\hsize}{!}{\includegraphics[angle=-90,width=16cm]{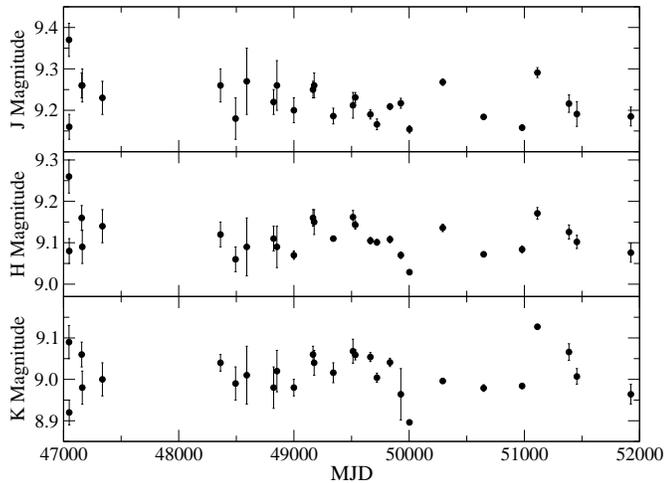}}
\caption{Infrared light curves of \object{BD~$+53\degr$2790}, taken during 1987\,--\,2001.}
\label{fig:irlc}
\end{figure}

Our infrared photometry coverage extends for $\approx 13$~yr and is
much more comprehensive than our optical photometry. The IR
long-term light curve is shown in Fig. \ref{fig:irlc}. Data have
been binned so that every point represents the average of all the
nights in a single run (excluding those with unacceptably large
photometric errors). 

%----------------------------------------------------------------------
%---------------------------------------------------------------------

As can be seen in Fig.~\ref{fig:irlc}, the range of variability is not very
large, with extreme values differing by $\approx 0.2\,{\rm mag}$ in all
three bands. Variability seems to be relatively random, in the sense
that there are no obvious long-term trends. The light curves for the three infrared
magnitudes are rather similar in shape, suggesting that the three
bands do not vary independently.

In spite of this, all colour-magnitude plots are dominated by scatter.
Moreover, an analysis of the temporal behaviour shows that there is no obvious 
pattern in the evolution of the source on the $H/(H-K)$ and $K/(H-K)$ planes, 
with frequent jumps between very distant points and no tendency to remain in any
particular region for any length of time. 

%--------------------------------------------------------------------
\begin{figure*} 
\centering
\resizebox{0.7\hsize}{!}{\includegraphics[angle=-90]{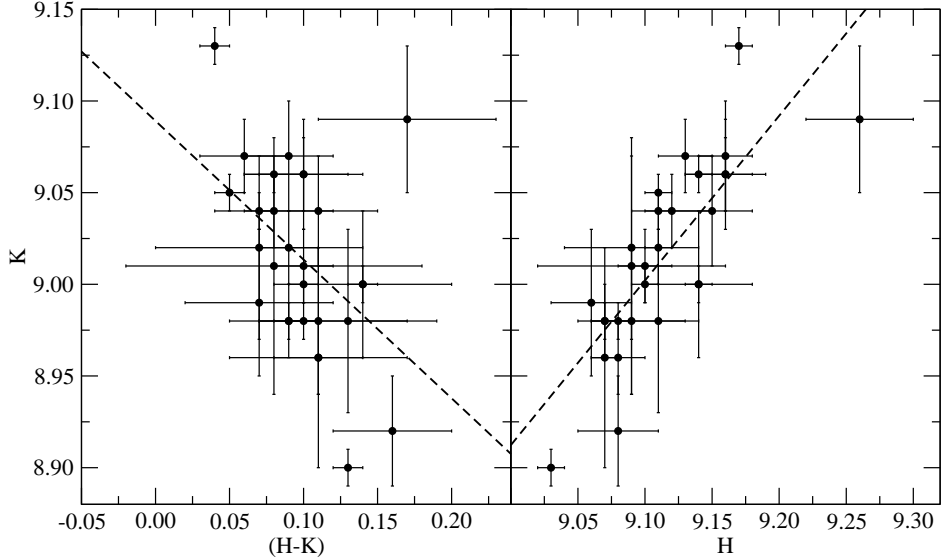}}
\caption{Colour-magnitude plots showing the evolution of the infrared
magnitudes. The strong correlation seen in the $K$/$(H-K)$ plane is not
a simple reflection of the fact that a brighter $K$ means a
smaller $(H-K)$, as the correlation between $H$ and $K$ is also
strong. Regression lines are shown as dashed lines. In the first case the correlation
coefficient is r$_{(H-K),K}$=-0.46 and the correlation is significant in a 98\% confidence level. In the latter 
case the correlatino coefficient is r$_{H,K}$=0.80 and the correlation is  also  significant at a 98\% confidence level.}
\label{fig:colmagplot}
\end{figure*}
%--------------------------------------------------------------------

The only plot in which a clear correlation stands out is the $K$/$(H-K)$
diagram (see Fig.~\ref{fig:colmagplot}). In principle, one would be 
tempted to dismiss this correlation as the simple reflection of stronger
variability in $K$ than in $H$, since $(H-K)$ would necessarily be
smaller for larger values of $K$. However a linear regression of $H$
against $K$ also shows a clear correlation, We find $a=0.89$,
$b=0.93$ and a correlation coefficient of $r^{2}=0.64$ for $K=aH+b$. Suspecting, then, 
that  linear correlation must be present in the $H$/$(H-K)$
plot as well, we also performed a linear regression.  In this case we found 
a very
poor correlation.

\begin{figure} 
\resizebox{\hsize}{!}{\includegraphics[angle=-90]{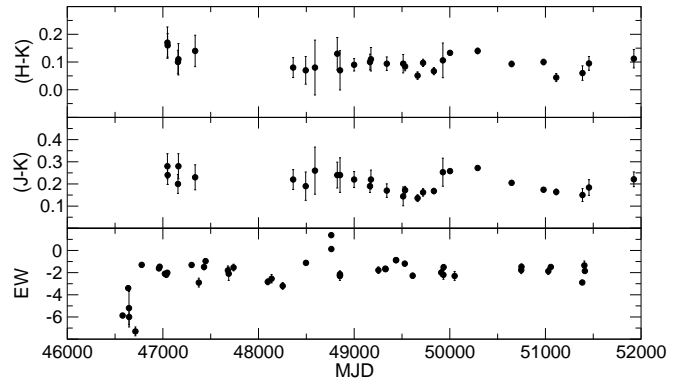}}
\vspace{0.3cm}
\caption{Evolution of the infrared colours in \object{BD~$+53\degr$2790}
  during 1987\,--\,1999 compared to that of the EW of H$\alpha$. Since
simultaneous measurements are rare, we cannot properly search for
correlations. The lack of any obvious correlated trends  could be
caused by the lack of long-term trends. }
\label{fig:ircol}
\end{figure}

Equally disappointing is the search for correlations between the 
EW of H$\alpha$ and the $(J-K)$ color. Even though our measurements
of these two quantities are not simultaneous, a look at their
respective evolutions (Fig.~\ref{fig:ircol}) shows no clear
correlation. 

%-----------------------------------------------------------------------
%------------------------------------------------------------------------

\subsection{Periodicity searches}

All the parameters of the H$\alpha$ emission line are clearly variable: EW,
FWHM, V/R ratio and peak separation. In the hope that the variation in
any of these parameters could give us information about the physical
processes causing them, we have searched the dataset for
periodicities. The large variety of resolutions, CCD configurations
and S/N ratios present in the data have hampered our attempts at a
homogeneous and coherent analysis. We have made an effort, within the
possibilities of the dataset, to use the same criteria to measure all
parameters on all spectra. We have used several different algorithms
(CLEAN, Scargle, PDM) in order to detect any obvious periodicities,
but with no success. No sign of any significant periodicity has been
found in any of the trials.

Likewise, we have explored possible periodicities in the infrared
light curves. While the $J$,
$H$ and $K$ magnitudes seem to vary randomly, we find a striking
apparent modulation of the $(J-K)$ colour.  Figure~\ref{fig:ircol}
shows an obvious trend in the evolution of $(J-K)$, with a suggestion
that the variability (with an amplitude $\sim 0.2$ mag) may be
(quasi-)periodic over a very long timescale, on the order of $\sim$5~y. Unfortunately, this
timescale is too long compared to our coverage to allow any certainty.

We have also folded the data using the period detected in the analysis
of the X-ray light curve of \object{4U~2206+54} (the presumably orbital 9.56-d
period, see Corbet \& Peele, 2001 and Rib\'o et al. 2005), without finding any significant periodic modulation.

\section{Intensive monitoring during the summer of 2000}

%-----------------------------------------------------------------------
\begin{figure} 
\centering
\resizebox{\hsize}{!}{\includegraphics[angle=-90]{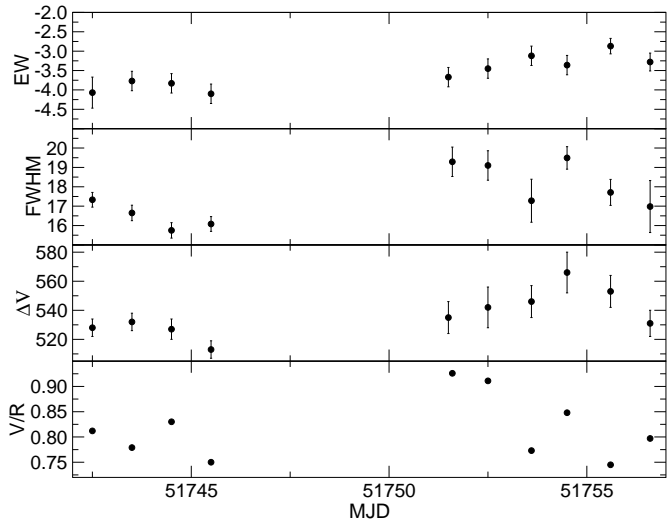}}
\caption{H$\alpha$ parameters -- EW (in \AA), FWHM (in \AA), peak separation 
(in km s$^{-1}$) and V/R ratio-- for the monitoring campaign in July
2000. There seems to be a high degree of correlation in the
evolution of EW, FWHM and peak separation, which is not shared by the
V/R ratio. }
\label{fig:july_par}
\end{figure}
%-----------------------------------------------------------------------

Considering the possibility that the lack of detectable periodicities
in our dataset was due to the varying resolutions and irregular time
coverage, during July 2000 we carried out more intensive spectroscopic
monitoring of \object{BD$\:+53^{\circ}\,$2790}. Observations were made from Skinakas
(Crete) and Loiano (Italy). We collected a set of 2 to 5 spectra per
night during two runs: from 17th to 20th July in Skinakas and from 26th to
31st July in Loiano.  The instrumental configurations were identical
to those described in Section 2. 

We fear that one of our objectives, the study of possible orbital
variations, may have been affected by an observational bias. The
presumed orbital period of the source is 9.56 days, probably too close
to the time lag (10 days) between the first observing night at Skinakas
and the first observing night at Loiano. Therefore we have not been
able to cover the whole orbital period. Indeed, the phases (in the
9.56~d cycle) at which the observations from Skinakas were taken, were
almost coincident with the phases during the first four Loiano
nights. For this reason, our coverage of the orbital period extends to
only $\approx60$\%, which is insufficient to effectively
detect any sort of modulation of any parameters at the orbital period.

Again, we have measured all parameters of the H$\alpha$ line, which
are shown in Fig~\ref{fig:july_par}. Contrary to what we saw when
considering the dataset for the 13 previous years, we find some degree of correlation
between EW, FWHM and $\Delta$V, while V/R seems to vary
independently. Since this correlation between the different line
parameters seems natural, we attribute the lack of correlations within
the larger dataset to the use of data of very uneven resolution and
quality. 

\begin{figure}
\centering 
\resizebox{0.7\hsize}{!}{\includegraphics[]{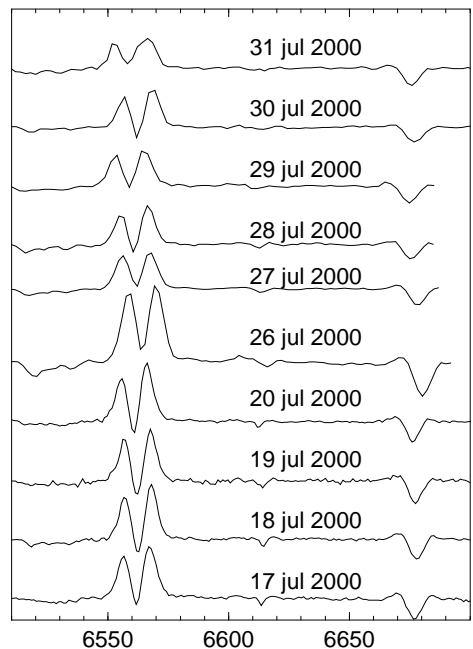}}
\caption{Evolution of H$\alpha$ line in \object{BD~$+53\degr$2790}
  during the monitoring campaign in July 2000. Note the moderate
night-to-night changes of the line profile and the important
difference between the spectra from the first and second week.}
\label{fig:july}
\end{figure}

We observe obvious changes in the depth of the central absorption core
in the H$\alpha$ line, which is seen sometimes reaching below the
continuum level, while in other occasions is above the
continuum (see Fig~\ref{fig:july}). Similar behaviour had already been observed
in 1986 (see
Fig.~\ref{fig:halpha}, but no further examples are found in our data
sample). Lines in the blue (3500--5500~\AA) are much more stable, as is also the case
when the longer term is considered. In this spectral range, the spectra resemble closely
those obtained at other epochs, with weak emission components
visible in \ion{He}{ii}~$\lambda$4686\AA\ and H$\beta$.

%------------------------------------------------------------------------

\section{Discussion}

\subsection{Reddening and distance to \object{BD~$+53\degr$2790}}

The reddening to \object{BD~$+53\degr$2790} can be estimated in
several different ways. Photometrically, from our value of
$E(b-y)=0.38\pm0.02$, using the correlation from \cite{sho83}, we
derive $E(B-V)=0.54\pm0.05$. An independent estimation can be made by
using the standard relations between the strength of Diffuse
Interstellar Bands (DIBs) in the spectra and reddening
(Herbig 1975). Using all the spectra obtained from the Cassini
telescope (for consistency), we derive $E(B-V)=0.57\pm0.06$ from the
$\lambda6613$\AA\ DIB and $E(B-V)=0.62\pm0.05$ from the
$\lambda4430$\AA\ DIB. All these values are consistent with each other,
therefore we take the photometric
value as representative of the reddening to  \object{BD~$+53\degr$2790}.

 From five $UBV$ measurements available in the
literature (including the one presented in this work), we find
$(B-V)=0.28\pm0.02$. With the $E(B-V)$ derived, this indicates an
intrinsic colour $(B-V)_{0}=-0.26\pm0.05$, typical of an early-type
star, confirming the validity of the reddening determination. As
discussed in NR01, the value of the absorption column derived from all X-ray 
observations is one order of magnitude larger than what is expected from the 
interstellar reddening. This  affirmation stands also when we consider the more 
accurate  measurement of  the absorption column 
(i.e., $\sim$1.0$\times$10$^{22}~cm^{-2}$) from {\it BeppoSax}
data (Torrej\'on et al, 2004; Masseti et al, 2004).

Averaging our 7 measurements of $y$ with the 5 $V$ measurements, we
find a mean value for \object{BD~$+53\degr$2790} of
$V=9.88\pm0.04$. Assuming a standard reddening law ($R=3.1$), we find
$V_{0}=8.21$. If the star has the typical luminosity of an O9.5V star
($M_{V}=-3.9$, see Martins et al. \cite{martins05}), then the distance to  \object{BD~$+53\degr$2790} is
$d\approx2.6$~kpc. This is closer than previous estimates (cf. NR01), because 
the absolute magnitudes of O-type stars have been lowered down in the most recent 
calibrations.

\subsection{Why \object{BD~$+53\degr$2790} is not a classical Be star}
Since its identification with 4U\,2206+54, \object{BD~$+53\degr$2790} has
always been considered a classical Be star, because of the presence of
shell-like emission lines in the red part of its spectrum. However,
the main observational
characteristics of \object{BD~$+53\degr$2790} differ considerably from
those of a classical Be star:

\begin{itemize}
\item The H$\alpha$ emission line presents a permanent (at least
  stable during 15 years) V$<$R asymmetry. Changes in the V/R ratio
  are not cyclical, as in classical Be stars undergoing V/R
  variability because of the presence of global one-armed oscillations
  (see \cite{oka00}). Moreover, the asymmetry survives large changes
  in all the other parameters of the emission line and is also present
  when there is basically no emission, which in a classical Be star
  would correspond to a disc-less state. This behaviour is
  fundamentally different of that seen in Be/X-ray binaries, where
  discs undergo processes of dispersion and reformation during which
  they develop instabilities that lead to long-term quasi-cyclical V/R
  variability (e.g., Negueruela et al. 2001 and Reig et al. 2000). 
\item In \object{BD~$+53\degr$2790} we observe strong night-to-night
  variability in both the shape and intensity of the H$\alpha$
  emission line. These variations affect both the strength of the 
  emission peaks and the depth of the central absorption 
  component. If the emission line did arise from an extended 
  quasi-Keplerian disc (as in Be stars), such variations would 
  imply global structural changes of the disc on timescales of
  a few hours and/or major changes in the intrinsic luminosity of the
  O star. Such behaviour is unprecedented in a Be star, where the
  circumstellar disc is believed to evolve on viscous timescales,
  on the order of months (Lee et al. \cite{lee91}; Porter \cite{porter99}).
\item Be stars display a clear correlation between the EW of
  H$\alpha$ and the infrared excess and between the infrared
  magnitudes and infrared colours, which reflect the fact that
  emission lines and infrared excess are produced in an envelope that
  adds its emission to that of the star, e.g., \cite{dw82}. 
  Such correlations are not readily detected  in \object{BD~$+53\degr$2790}. 
  The evolution of observables (both IR magnitudes and H$\alpha$
  line parameters) lacks any clear long-term trends. The star's 
  properties may be described to be highly variable on short timescales
  and very stable on longer timescales, without obvious long-term
  variations (except for, perhaps, the $(J-K)$ colour).
\item Photometrically, Be/X-ray systems
  are characterised by large variations in both magnitudes and to a
  lesser extent in colour (e.g, Negueruela et al. 2001; Clark et al. 1999 and Clark et al. 2001b),
  associated with the periods of structural changes in their circumstellar
  discs. In contrast, the magnitudes and colours of \object{BD
  $+53\degr$2790} remain basically stable, with small random
  fluctuations, as is typical of isolated O-type stars.
\end{itemize}
  
As a matter of fact, the only High-Mass X-ray Binary presenting some
similarities to \object{BD~$+53\degr$2790} in its photometric
behaviour is \object{LS~5039}/\object{RX~J1826.2$-$1450}. As
\object{BD~$+53\degr$2790}, it displays little variability in $UBV$
and moderate variability in the infrared magnitudes, see
\cite{jsc01a}. \object{RX~J1826.2$-$1450} is believed to be, like
\object{4U~2206+54}, powered by accretion from the wind of a main-sequence 
O-type star; see \cite{msg02}, Rib\'o et al (1999) and Reig et al. (2003).

\subsection{What is \object{BD~$+53\degr$2790}?}

We estimate that the most likely spectral classification of \object{BD~$+53\degr$2790}
is O9.5Vp. However some remarkable peculiarities have been noticed:
while the blue spectrum of \object{BD~$+53\degr$2790} suggests an 09.5 spectral
type, there are a few metallic lines reminiscent of a later-type spectrum (see NR01); the UV lines
support the main sequence luminosity classification, but the Paschen lines resemble
those of a supergiant. 

\begin{figure*}
\centering
\resizebox{0.7\hsize}{!}{\includegraphics{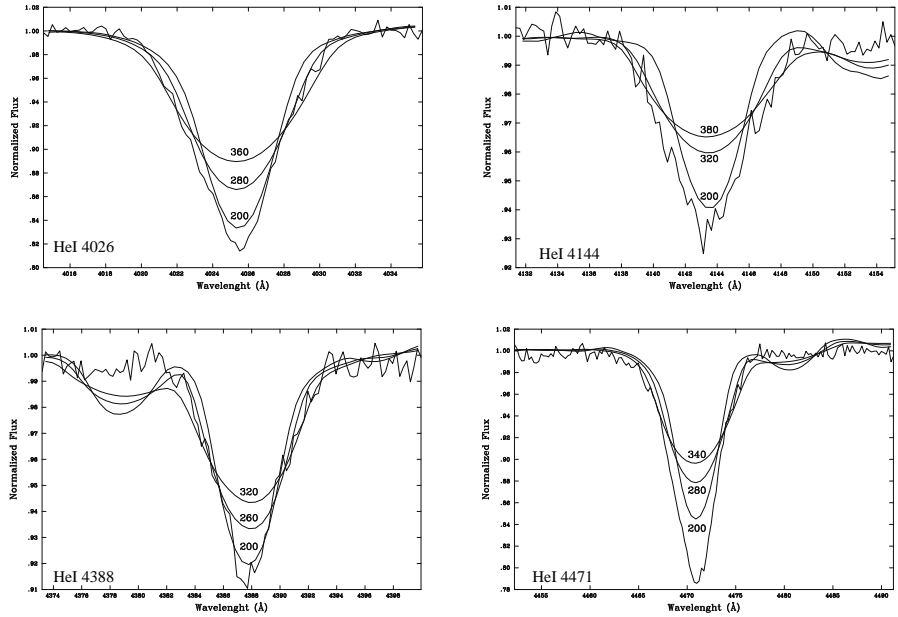}}
\caption{Normalised profiles of selected \ion{He}{i} lines (namely,\ion{He}{i}~$\lambda$4026, $\lambda$4144,
$\lambda$4388, and $\lambda$4471~\AA) from \object{BD~$+53\degr$2790} together with those of the same lines from
\object{10~Lac} but artificially broadened to 200~km~s$^{-1}$ and to those rotational
velocities yielding upper and lower envelopes to the with of the  \object{BD~$+53\degr$2790}
lines (their values are shown at the peak of the profile, in units of km~s$^{-1}$). In all cases rotational 
velocities above 200~km~s$^{-1}$ are needed to reproduce the line widths.}
\label{fig:rotation}
\end{figure*}

In order to obtain a measure of the rotational velocity of \object{BD~$+53\degr$2790} we 
have created a grid of artificially rotationally broadened spectra from that of the
standard O9V star 10 Lac. We have chosen 10 Lac because of its very low projected rotational
velocity and because the spectrum of \object{BD~$+53\degr$2790} is close to that of
a O9V star.
In Fig. \ref{fig:rotation} normalised profiles of a set of
selected helium lines (namely, \ion{He}{i}~$\lambda$4026, $\lambda$4144,
$\lambda$4388, and $\lambda$4471~\AA) are shown together with the artificially
broadened profile of \object{10~Lac}, at 200~km~s$^{-1}$
and those rotational velocities producing upper and lower envelopes to the
widths of the observed profiles of \object{BD~$+53\degr$2790}. The rotational
velocity of \object{BD~$+53\degr$2790} must be above 200~km~s$^{-1}$. For
each line, the average of the rotational velocities yielding the upper
and lower envelopes were taken as a representative measurement of the rotational
velocity derived from that line. The results of these measurements are summarised
in Table \ref{tab:rotation}. We estimated the averaged rotational velocity 
of \object{BD~$+53\degr$2790} to be 315$\pm$70~km~s$^{-1}$.

%-----------------------------------------------------------------------
%-----------------------------------------------------------------------

%-----------------------------------------------------------------
\begin{table}
\caption{Summary of the measured rotational velocities for the selected
helium lines shown in Fig. \ref{fig:rotation}.}
\label{tab:rotation}
\centering
\begin{tabular}{ccc}
\hline
\hline
Line & Rot. Vel. & Average \\
(\AA) & (km~s$^{-1}$) & (km~s$^{-1}$) \\
\hline
\multirow{2}*{\ion{He}{i}$\lambda$4026} & 280 &  \multirow{2}*{320$\pm$40}\\
		 & 360 &	    \\
\hline
\multirow{2}*{\ion{He}{i}$\lambda$4144} & 320 &  \multirow{2}*{350$\pm$30}\\
		 & 380 &	    \\
\hline
\multirow{2}*{\ion{He}{i}$\lambda$4388} & 260 &  \multirow{2}*{290$\pm$30}\\
		 & 320 &	    \\
\hline
\multirow{2}*{\ion{He}{i}$\lambda$4471} & 260 &  \multirow{2}*{300$\pm$40} \\
                 & 340 &             \\
\hline
\end{tabular}
\end{table}
%-----------------------------------------------------------------

%Due to the closeness of this binary system, we could expect that 
%the rotational and orbital motions of the optical companion could be 
%synchronised. If that is the case for \object{BD~$+53\degr$2790}, considering 
%a stellar radius of 7.18~$R_{\sun}$ (expected value for an O9.5V star, see Martins et al. \cite{martins05})
%and the orbital period of the system ($\sim$9.6 days, see Corbet \& Peele, 2001, and 
%Rib\'o et al. 2005), we derive an expected rotational velocity
% on the order of $\sim$40~km~s$^{-1}$. This number is incompatible with a
%projected rotational velocity of $\sim$300~km~s$^{-1}$. 
%According to this estimate, the system would not be very evolved.

Comparison of the helium profiles with those rotationally broadened from \object{10~Lac}
shows that the observed helium profiles in \object{BD~$+53\degr$2790} are stronger than
what is expected for a normal O9.5V star.
The strength of the He lines suggests the possibility that \object{BD~$+53\degr$2790}  may be
related to the He-strong stars. These are a small group 
of stars, with spectral types clustering around B2~V, that show anomalously strong helium lines.
A well known O-type star believed to be related to He-strong stars is
$\theta^1$ Ori C, which is known to vary in spectral type from  O6 to O7 
(Donati et al. \cite{donati02}, Smith \& Fullerton \cite{smith05}). \object{BD~$+53\degr$2790} 
could be the second representative of this class of objects among O-type stars. 
He-strong stars display a remarkable set of peculiarities: 
oblique dipolar magnetic fields,  magnetically controlled winds, and chemical surface anomalies,
among others. Usually these stars are distributed along the ZAMS 
(Pedersen \& Thomsen, 1997; Walborn, 1982; Bohlender et al. 1987; Smith \& Groote 2001).

% This fact opens the possibility that 
%\object{BD~$+53\degr$2790} is the second found early type representative of the well known
%class of He-strong stars (being $\theta^1$ Ori C the first one, see Donati et al. 2002 and Smith
%\& Fullerton 2005). These are a small group 
%of stars, with spectral types gathered around B2~V, that show anomalously
%strong helium lines. This is a type of stars with a remarkable set of peculiarities: 
%oblique dipolar magnetic fields,  magnetically controlled winds, and chemical surface anomalies,
%among others. Usually this kind of stars are distributed along the ZAMS 
%(Pedersen \& Thomsen, 1997; Walborn, 1982; Bohlender et al. 1987; Smith \& Groote 2001).

A rich variety of phenomena have been observed in these objects: in the UV, they can show
red shifted emission of the \ion{C}{iv} and \ion{Si}{iv} resonance lines  (sometimes variable);
in the optical bands they are characterized by periodically modulated H$\alpha$ emission, 
high level Balmer lines appearing at certain rotational phases and periodically modulated 
variability in He lines, sometimes showing  emission at \ion{He}{ii}~$\lambda$4686~\AA. They
can also show photometric variability with eclipse-like light curves.

Except for the periodic modulation of the variations, \object{BD~$+53\degr$2790} shares
many of these peculiarities. In particular, together with the apparent high helium
abundance, \object{BD~$+53\degr$2790} shows variable H$\alpha$ emission and
\ion{He}{ii}~$\lambda$4686~\AA~emission, the UV spectrum shows apparently prominent
P-Cygni profiles at \ion{C}{iv} and \ion{Si}{iv} resonance lines (see NR01). 
In contrast a wind slower than expected is found (see Rib\'o et al. 2005), which can be an indication 
of some red-shifted excess of emission in these lines. In He-strong stars the wind is
conducted along the magnetic field lines into a torus-like envelope located at the magnetic 
equator. This configuration can lead to the presence of double emission peaks in H$\alpha$, which
resemble those seen in \object{BD~$+53\degr$2790}, but which usually show a modulation on
the rotational period.
The complexity and shape of the double peak will depend on the angle between magnetic 
and rotational axes and the line of sight to the observer (see Townsend et al. 2005).

A rotationally dominated circumstellar envelope is clearly present in \object{BD~$+53\degr$2790}, as indicated by the infrared magnitudes, 
the emission in Balmer and some helium lines and the correlations between H$\alpha$ line parameters. 
However the structure of this circumstelar envelope  clearly differs from those seen in Be stars. 
Following the analogy with He-strong stars, the existence of a circumstellar disk-like structure is also common to 
these type of objects also. The only difficulty to accept \object{BD+53$^\circ$2790} as a He-strong star 
 is the apparent lack  of rotational modulation of the emission lines parameters. Given the rotational 
 velocities derived, we could expect a rotational period of a few days. In addition to the problems
in the diverse origin of our data (see section \ref{baddata}), the sampling on time of our measurements is not adequate to find
variations on time scales of a few days (modulated with the rotational period), thus we cannot discard 
yet the presence of rotational periodicity.  The idea of a magnetically driven wind contributing to a 
dense disk-like structure is not strange even in the modelling of Be stars' circumstellar envelopes. The wind 
compressed disk of Bjorkman \& Cassinelli (\cite{bjorkman92}) was shown to be compatible with observations 
only if a magnetic field  on the order of tens of Gauss was driving the wind from the polar 
caps onto the equatorial zone (Porter \cite{porter97}).

A careful inspection to the correlation seen in Fig. \ref{fig:halpha_vs_hei} between the
 He\,{\sc i}~$\lambda$\,6678 and H$\alpha$ EWs shows that there is a common
component to the emission of both lines. H$\alpha$ emission, then, will have at least two contributions:
a P-Cygni like contribution (as seen in the 1992 spectra, see Fig. \ref{fig:halpha}, where the double peak
structure disappears and only the red peak survives) and an additional variable double
peaked structure. The relative variation of both components may hide any
periodic modulation present.

Therefore,
we can conclude that this is a very peculiar O9.5V star where most likely a global strong
magnetic field may be responsible for most of the behaviour seen so far.

\section{Conclusion}

We have presented the results of $\sim$14 years of spectroscopic 
and optical/infrared photometric monitoring  of
\object{BD~$+53\degr$2790}, the optical component of the Be/X-ray binary \object{4U\,2206+54}. 
The absence of any obvious long-term trends in the evolution of different parameters and
fundamentally the absence of correlation between the EW  of H$\alpha$ and the infrared 
magnitudes and associated colours makes untenable a Be classification for the star. Based on a careful inspection
to the source spectrum in the classification region and the peculiar behavior of the H$\alpha$ emisson line, we conclude 
that the object is likely to be a single peculiar O-type star (O9.5Vp) and an early-type analogue 
to He-strong stars.

\acknowledgements

We would like to thank the UK PATT and the Spanish CAT panel for supporting 
our long-term monitoring campaign. We are grateful to the INT
service programme for additional optical observations. The
1.5-m TCS is operated by the Instituto de Astrof\'{\i}sica de Canarias at the
Teide Observatory, Tenerife. The 
JKT and INT are operated on the island of La Palma by the Royal
Greenwich Observatory in the Spanish Observatorio del Roque de
Los Muchachos of the Instituto de Astrof\'{\i}sica de Canarias. The
1.5-m telescope at Mount Palomar is jointly owned by the California
Institute of Technology and the Carnegie Institute of Washington. 
The G.~D.~Cassini telescope is operated at the Loiano Observatory by the 
Osservatorio Astronomico di Bologna. 
Skinakas Observatory is a collaborative project of the University of Crete, 
the Foundation for Research and Technology-Hellas and the Max-Planck-Institut 
f\"ur Extraterrestrische Physik.

This  research has made use of the Simbad database, operated at CDS, 
Strasbourg (France), and of the La Palma  Data Archive. Special thanks 
to Dr. Eduard Zuiderwijk for his help  with the archival data. 

We are very grateful to the many astronomers who have taken part in
observations for this campaign. In particular, Chris 
Everall obtained and reduced the $K$-band spectroscopy and Miguel \'Angel
Alcaide reduced most of the H$\alpha$ spectra. 

P.B. acknowledges support by the Spanish Ministerio de Educaci\'on y Ciencia 
through grant ESP-2002-04124-C03-02. I.N. is a researcher of the programme {\em Ram\'on y Cajal}, 
funded by the Spanish Ministerio de Educaci\'on y Ciencia and the University of Alicante, with partial 
support from the Generalitat Valenciana and the European Regional Development Fund (ERDF/FEDER).
 This research is partially supported by the Spanish MEC through grants
AYA2002-00814 and ESP-2002-04124-C03-03.

%\Online

%------------------------------------------------------------------------
\begin{table*}
\caption{Log of spectroscopic observations during 2000.}
\label{tab:log2000}
\begin{center}
\begin{tabular}{lcccc}
\hline\hline
Date         & Tel & Configuration & Detector & $\lambda$ Range (\AA) \\
\hline
Jul 17, 2000 & SKI       & grating 1302 l/mm blazed at 5500\AA  & SITe  & 5520 - 7560 \\
Jul 18, 2000 & SKI       & grating 1302 l/mm blazed at 5500\AA & SITe  & 5520 - 7560 \\
Jul 19, 2000 & SKI       & grating 1302 l/mm blazed at 5500\AA & SITe  & 5520 - 7560 \\
Jul 20, 2000 & SKI       & grating 1302 l/mm blazed at 5500\AA & SITe  & 5520 - 7560 \\
Jul 21, 2000 & SKI       & grating 1302 l/mm blazed at 4800\AA  & SITe  & 3800 - 5700 \\
Jul 22, 2000 & SKI       & grating 1302 l/mm blazed at 4800\AA  & SITe  & 3800 - 5700 \\
Jul 25, 2000 & BOL       & BFOSC + gr\#6  & EEV & 3100 - 5300  \\
Jul 25, 2000 & BOL       & BFOSC + gr\#7  & EEV & 4200 - 6700  \\
Jul 25, 2000 & BOL       & BFOSC + gr\#8  & EEV & 6100 - 8200  \\
Jul 26, 2000 & BOL       & BFOSC + gr\#6  & EEV & 3100 - 5300  \\
Jul 26, 2000 & BOL       & BFOSC + gr\#7  & EEV & 4200 - 6700  \\
Jul 26, 2000 & BOL       & BFOSC + gr\#8  & EEV & 6100 - 8200  \\
Jul 27, 2000 & BOL       & BFOSC + gr\#6  & EEV & 3100 - 5300  \\
Jul 27, 2000 & BOL       & BFOSC + gr\#7  & EEV & 4200 - 6700  \\
Jul 27, 2000 & BOL       & BFOSC + gr\#8  & EEV & 6100 - 8200  \\
Jul 28, 2000 & BOL       & BFOSC + gr\#6  & EEV & 3100 - 5300  \\
Jul 28, 2000 & BOL       & BFOSC + gr\#7  & EEV & 4200 - 6700  \\
Jul 28, 2000 & BOL       & BFOSC + gr\#8  & EEV & 6100 - 8200  \\
Jul 29, 2000 & BOL       & BFOSC + gr\#6  & EEV & 3100 - 5300  \\
Jul 29, 2000 & BOL       & BFOSC + gr\#8  & EEV & 6100 - 8200  \\
Jul 29, 2000 & BOL       & BFOSC + gr\#9+\#10  & EEV & 3750 - 8000  \\
Jul 30, 2000 & BOL       & BFOSC + gr\#8  & EEV & 6100 - 8200  \\
Jul 30, 2000 & BOL       & BFOSC + gr\#9+\#10  & EEV & 3750 - 8000  \\
Oct 05, 2000 & SKI       & grating 1302 l/mm blazed at 5500\AA  & SITe  & 5520 - 7560 \\
Oct 16, 2000 & SKI       & grating 1302 l/mm blazed at 5500\AA  & SITe  & 5520 - 7560 \\
Oct 17, 2000 & SKI       & grating 1302 l/mm blazed at 4800\AA  & SITe  & 3800 - 5700 \\
\hline
\end{tabular}
\end{center}
\end{table*}
%-----------------------------------------------------------------------

%---------------------------------------------------------------------------------
\begin{center}
\begin{longtable}{l c c c c c l}
\caption{Log of spectroscopic observations. Some representative spectra are
displayed in Fig. \ref{fig:halpha} (marked with '*')}
\label{tab:log} \\
 \hline\hline
Date         & Tel & Configuration & Detector & $\lambda$ Range &  EW of H$\alpha$ \\
             &     &               &          &   (\AA)         &    (\AA)     \\
\hline\hline
\endfirsthead
\multicolumn{7}{c}{Table \ref{tab:log}: Continued.}\vspace{0.35cm} \\
\hline \hline
Date         & Tel & Configuration & Detector & $\lambda$ Range &  EW of H$\alpha$ \\
             &     &               &          &   (\AA)         &    (\AA)     \\
\hline
\endhead
\hline
\endlastfoot
\hline
\multicolumn{7}{l}{\emph{Continued...}}
\endfoot
May 28, 1986(*)& INT	   & IDS + 500 mm & GEC1 & 6450 $-$ 6830 &  $-5.87\pm0.17$ \\
Jul 26, 1986(*)& INT	   & IDS + 500 mm & GEC1 & 6495 $-$ 6695 &  $-3.40\pm0.20$ \\
Aug 03, 1986   & INT	   & IDS + 235 mm & IPCS & 6010 $-$ 7020 &  $-5.20\pm1.50$ \\
Aug 04, 1986   & INT	   & IDS + 235 mm & IPCS & 6010 $-$ 7020 &  $-6.00\pm0.90$ \\
Sep 07, 1986   & INT	   & IDS + 235 mm & GEC1 & 4000 $-$ 8000 &		   \\
Oct 09, 1986(*)& INT	   & IDS + 500 mm & GEC1 & 6465 $-$ 6665 &  $-7.30\pm0.40$ \\
Dec 24, 1986(*)& INT	   & IDS + 500 mm & GEC1 & 6250 $-$ 6875 &  $-1.30\pm0.11$ \\
Jun 12, 1987   & INT	   & IDS + 235 mm & GEC1 & 6330 $-$ 6770 &  $-1.62\pm0.18$ \\
Jun 20, 1987   & INT	   & IDS + 235 mm & GEC1 & 6080 $-$ 6900 &  $-1.47\pm0.21$ \\
Aug 13, 1987(*)& INT	   & IDS + 500 mm & GEC1 & 6375 $-$ 6765 &  $-2.10\pm0.21$ \\
Aug 28, 1987   & INT	   & IDS + 235 mm & GEC1 & 6340 $-$ 6770 &  $-2.19\pm0.22$ \\
Sep 08, 1987   & INT	   & IDS + 500 mm & GEC1 & 6340 $-$ 6730 &  $-2.01\pm0.12$ \\
May 19, 1988   & INT	   & IDS + 500 mm & IPCS & 6240 $-$ 6720 &  $-1.31\pm0.19$ \\
Aug 02, 1988(*)& INT	   & IDS + 235 mm & GEC4 & 6230 $-$ 6860 &  $-2.90\pm0.40$ \\
Sep 26, 1988   & INT	   & IDS + 500 mm & GEC4 & 6455 $-$ 6655 &  $-1.49\pm0.25$ \\
Oct 12, 1988   & INT	   & IDS + 500 mm & GEC4 & 6325 $-$ 6950 &  $-0.96\pm0.17$ \\
Jun 02, 1989   & INT	   & IDS + 235 mm & IPCS & 6230 $-$ 6875 &  $-1.80\pm0.40$ \\
Jun 11, 1989   & INT	   & IDS + 235 mm & IPCS & 5970 $-$ 7010 &  $-2.10\pm0.60$ \\
Jul 31, 1989(*)& INT	   & IDS + 235 mm & IPCS & 6205 $-$ 6870 &  $-1.54\pm0.30$ \\
Jul 24, 1990   & INT	   & IDS + 235 mm & GEC6 & 6350 $-$ 6780 &  $-2.83\pm0.26$\\
Sep 02, 1990   & INT	   & IDS + 235 mm & GEC6 & 6345 $-$ 6775 &  $-2.55\pm0.38$ \\
Dec 27, 1990   & INT	   & IDS + 500 mm & GEC6 & 6470 $-$ 6670 &  $-3.20\pm0.30$ \\
Aug 28, 1991(*)& INT	   & IDS + 500 mm & GEC6 & 6480 $-$ 6680 &  $-1.12\pm0.18$ \\
May 18, 1992(*)& PAL	   & f/8.75 Cass  & CCD9 & 6255 $-$ 6938 &  $+1.38\pm0.13$ \\
May 19, 1992   & PAL	   & f/8.75 Cass &  CCD9 & 6522 $-$ 6663 &  $+0.13\pm0.05$ \\
Aug 16, 1992(*)& PAL	   & f/8.75 Cass &  CCD9 & 6255 $-$ 6930 &  $-2.30\pm0.40$ \\
Aug 17, 1992   & PAL	   & f/8.75 Cass &  CCD9 & 6255 $-$ 6930 &  $-2.14\pm0.30$ \\
Aug 18, 1992   & PAL	   & f/8.75 Cass &  CCD9 & 6255 $-$ 6930 &  $-2.23\pm0.30$ \\
Sep 23, 1993(*)& PAL	   & f/8.75 Cass &  CCD9 & 6280 $-$ 6960 &  $-1.78\pm0.30$ \\
Dec 05, 1993   & PAL	   & f/8.75 Cass &  CCD9 & 6259 $-$ 6936 &  $-1.64\pm0.19$ \\
Dec 06, 1993   & PAL	   & f/8.75 Cass &  CCD9 & 4300 $-$ 5000 &		   \\ 
Dec 07, 1993   & PAL	   & f/8.75 Cass &  CCD9 & 6260 $-$ 6940 &  $-1.69\pm0.15$ \\ 
Mar 26, 1994   & JKT	   & RBS  & EEV7 & 5700 $-$ 6710 & $-0.88\pm0.18$ \\
Mar 27, 1994   & JKT	   & RBS  & EEV7 & 5700 $-$ 6710 & $-0.87\pm0.15$ \\
Jun 25, 1994   & JKT	   & RBS  & EEV7 & 4200 $-$ 5200 &		   \\
Jun 26, 1994   & JKT	   & RBS  & EEV7 & 6070 $-$ 7040 & $-1.19\pm0.21$ \\
Jun 27, 1994   & JKT	   & RBS  & EEV7 & 4200 $-$ 5200 &		  \\
Sep 16, 1994(*)& JKT	   & RBS  & EEV7 & 5825 $-$ 6890 & $-2.28\pm0.24$ \\
Sep 16, 1994   & JKT	   & RBS  & EEV7 & 8100 $-$ 9100 &		  \\	    
Sep 17, 1994   & JKT	   & RBS  & EEV7 & 3900 $-$ 4950 &		  \\	    
Jul 11, 1995   & INT	   & IDS + 235 mm & TEK3 & 4080 $-$ 4940 &		   \\ 
Jul 12, 1995(*)& INT	   & IDS + 235 mm & TEK3 & 6430 $-$ 7286 & $-2.00\pm0.30$ \\
Aug 04, 1995   & JKT	   & RBS  & TEK4 & 6360 $-$ 7265 &	    $-2.20\pm0.40$ \\
Aug 04, 1995   & JKT	   & RBS  & TEK4 & 8200 $-$ 9000 &		  \\	    
Aug 05, 1995   & JKT	   & RBS  & TEK4 & 4100 $-$ 5050 &		  \\	    
Aug 06, 1995   & JKT	   & RBS  & TEK4 & 6420 $-$ 6755 & $-1.51\pm0.25$ \\  
Aug 07, 1995   & JKT	   & RBS  & TEK4 & 4000 $-$ 4450 &		   \\	    
Sep 22, 1995   & CRAO	   & Coude  & EEV15$-$11 & 4400 $-$ 4950 &		    \\
Nov 29, 1995   & JKT	   & RBS  & TEK4 & 6100 $-$ 6900 &	     $-2.30\pm0.40$ \\       
Jun 30, 1997   & CRAO	   & Coude  & EEV15$-$11 & 4200 $-$ 5100 &		   \\ 
Oct 26, 1997   & JKT	   & RBS    & TEK4	 & 5904 $-$ 6818 &   $-1.77\pm0.32$ \\        
Oct 27, 1997   & JKT	   & RBS  & TEK4 & 8200 $-$ 9000 &		  \\	     
Oct 28, 1997(*)& JKT	   & RBS  & TEK4 & 6380 $-$ 6720 &   $-1.47\pm0.25$ \\
Aug 03, 1998   & INT	   & IDS + 235 mm & EEV42& 3700 $-$ 5050 &		    \\  	   
Aug 04, 1998(*)& INT	   & IDS + 235 mm & EEV42& 5800 $-$ 7100 & $-1.87\pm0.30$ \\		      
Aug 31, 1998   & CRAO	   & Coude  & EEV15$-$11 & 6530 $-$ 6600 &  $-1.49\pm0.02$ \\ 
Aug 31, 1998   & CRAO	   & Coude  & EEV15$-$11 & 6645 $-$ 6620 &		   \\		  
Jul 26, 1999   & SKI	   & 1201~line~mm$^{-1}$ grating & ISA SITe CCD & 5520 $-$ 7560 &   $-2.89\pm0.07$ \\		       
Aug 17, 1999(*)& BOL	   & BFOSC + gr\#9+\#12  & Loral & 5300 $-$ 9000 &  $-1.35\pm0.40$ \\ 
Aug 17, 1999   & BOL	   & BFOSC + gr\#7  & Loral & 4200 $-$ 6700  &  	      \\	  
Aug 22, 1999   & BOL	   & BFOSC + gr\#8  & Loral & 6100 $-$ 8200  &  $-1.85\pm0.10$ \\ 
Aug 22, 1999   & BOL	   & BFOSC + gr\#7  & Loral & 4200 $-$ 6700  &  	       \\
\end{longtable}
\end{center}
%---------------------------------------------------------------------------------

%------------------------------------------------------------
\begin{center}
\begin{longtable}{l c c c c c l}
\caption{Observational details, IR photometry.}\\
\label{tab:observations} \\
\hline\hline
Date & MJD & J & H & K & L$^{\prime}$ & Telescope  \\ 
& & & & & & \\
\hline\hline
\endfirsthead
\multicolumn{7}{c}{Table \ref{tab:observations}: Continued.}\vspace{0.35cm} \\
\hline \hline
Date & MJD & J & H & K & L$^{\prime}$ & Telescope  \\
& & & & & & \\
\hline
\endhead
\hline
\endlastfoot
\hline
\multicolumn{7}{l}{\emph{Continued...}}
\endfoot

Sep 06, 1987 & 47044.5   &  9.37$\pm$0.03 &  9.26$\pm$0.03 &  9.09$\pm$0.03 & 9.0$\pm$0.4 & TCS \\
Sep 11, 1987 & 47049.5   &  9.70$\pm$0.04 &  9.48$\pm$0.04 &  9.28$\pm$0.04 & & TCS \\
Nov 24, 1987 & 47123.5   &  9.16$\pm$0.03 &  9.08$\pm$0.03 &  8.92$\pm$0.03 & 8.71$\pm$0.06 & UKIRT \\
Dec 28, 1987 & 47157.5   &  9.26$\pm$0.03 &  9.16$\pm$0.03 &  9.06$\pm$0.03 & & TCS \\
Jan 01, 1988 & 47161.5   &  9.28$\pm$0.04 &  9.10$\pm$0.04 &  8.99$\pm$0.04 & 8.8$\pm$0.4 & TCS \\
Jan 03, 1988 & 47163.5   &  9.24$\pm$0.03 &  9.08$\pm$0.03 &  8.97$\pm$0.03 & 8.8$\pm$0.3 & TCS \\
Jun 23, 1988 & 47335.5   &  9.28$\pm$0.04 &  9.17$\pm$0.04 &  9.01$\pm$0.04 & & TCS \\
Jun 24, 1988 & 47336.5   &  9.25$\pm$0.03 &  9.15$\pm$0.03 &  8.98$\pm$0.03 & & TCS \\
Jun 28, 1988 & 47340.5   &  9.16$\pm$0.04 &  9.11$\pm$0.04 &  9.00$\pm$0.04 & & TCS \\
Apr 14, 1991 & 48360.5   &  9.25$\pm$0.02 &  9.12$\pm$0.03 &  9.05$\pm$0.02 & & TCS \\
Apr 17, 1991 & 48363.5   &  9.26$\pm$0.06 &  9.12$\pm$0.03 &  9.03$\pm$0.02 & & TCS \\
Aug 23, 1991 & 48491.5   &  9.22$\pm$0.03 &  9.08$\pm$0.02 &  9.01$\pm$0.02 & & TCS \\
Aug 24, 1991 & 48492.5   &  9.16$\pm$0.09 &  9.04$\pm$0.05 &  9.00$\pm$0.05 & & TCS \\
Aug 25, 1991 & 48493.5   &  9.18$\pm$0.04 &  9.05$\pm$0.03 &  8.95$\pm$0.03 & & TCS \\
Aug 27, 1991 & 48495.5   &  9.17$\pm$0.03 &  9.06$\pm$0.02 &  9.00$\pm$0.03 & & TCS \\
Aug 28, 1991 & 48496.5   &  9.19$\pm$0.04 &  9.05$\pm$0.03 &  8.98$\pm$0.05 & & TCS \\
Nov 29, 1991 & 48589.5   &  9.30$\pm$0.10 &  9.10$\pm$0.10 &  9.00$\pm$0.10 & & TCS \\
Dec 01, 1991 & 48591.5   &  9.24$\pm$0.05 &  9.08$\pm$0.04 &  9.01$\pm$0.04 & & TCS \\
Jul 20, 1992 & 48823.5   &  9.22$\pm$0.03 &  9.11$\pm$0.03 &  8.98$\pm$0.05 & & TCS \\
Aug 04, 1992 & 48838.5   &  9.33$\pm$0.09 &  9.09$\pm$0.04 &  8.93$\pm$0.05 & & TCS \\
Aug 21, 1992 & 48855.5   &  9.29$\pm$0.04 &  9.07$\pm$0.03 &  9.03$\pm$0.06 & & TCS \\
Aug 21, 1992 & 48855.5   &  9.25$\pm$0.05 &  9.07$\pm$0.03 &  8.94$\pm$0.06 & & TCS \\
Aug 22, 1992 & 48856.5   &  9.14$\pm$0.06 &  9.04$\pm$0.04 &  9.07$\pm$0.04 & & TCS \\
Aug 23, 1992 & 48857.5   &  9.29$\pm$0.07 &  9.16$\pm$0.09 &  9.11$\pm$0.06 & & TCS \\
Jan 12, 1993 & 48999.5   &  9.20$\pm$0.03 &  9.07$\pm$0.01 &  8.98$\pm$0.02 & & TCS \\
Jun 22, 1993 & 49160.5   &  9.23$\pm$0.01 &  9.17$\pm$0.01 &  9.09$\pm$0.01 & 8.91$\pm$0.03 & UKIRT\\
Jun 30, 1993 & 49169.5   &  9.27$\pm$0.03 &  9.14$\pm$0.02 &  9.02$\pm$0.02 & 8.9$\pm$0.3 & UKIRT\\
Jul 06, 1993 & 49175.5   &  9.26$\pm$0.02 &  9.14$\pm$0.03 &  9.03$\pm$0.02 & 8.9$\pm$0.2 & UKIRT\\
Jul 07, 1993 & 49176.5   &  9.26$\pm$0.03 &  9.15$\pm$0.02 &  9.05$\pm$0.03 & & UKIRT\\
Dec 20, 1993 & 49341.5   &  9.11$\pm$0.01 &  9.11$\pm$0.01 &  8.95$\pm$0.02 & & TCS  \\
Dec 20, 1993 & 49341.5   &  9.11$\pm$0.01 &  9.08$\pm$0.01 &  8.93$\pm$0.01 & & TCS  \\
Dec 20, 1993 & 49341.5   &  9.18$\pm$0.03 &  9.06$\pm$0.01 &  8.99$\pm$0.06 & & TCS  \\
Dec 21, 1993 & 49342.5   &  9.26$\pm$0.02 &  9.18$\pm$0.01 &  9.09$\pm$0.03 & & TCS  \\
Dec 21, 1993 & 49342.5   &  9.27$\pm$0.03 &  9.13$\pm$0.01 &  9.13$\pm$0.01 & & TCS  \\
Jun 08, 1994 & 49511.5   &  9.16$\pm$0.02 &  9.16$\pm$0.01 &  9.01$\pm$0.01 & & TCS  \\
Jun 09, 1994 & 49512.5   &  9.28$\pm$0.01 &  9.13$\pm$0.02 &  9.06$\pm$0.03 & & TCS  \\
Jun 10, 1994 & 49513.5   &  9.16$\pm$0.03 &  9.07$\pm$0.01 &  9.04$\pm$0.03 & & TCS  \\
Jun 11, 1994 & 49514.5   &  9.24$\pm$0.06 &  9.28$\pm$0.04 &  9.17$\pm$0.05 & & TCS \\
Jun 21, 1994 & 49524.5   &  9.22$\pm$0.01 &  9.12$\pm$0.01 &  9.03$\pm$0.01 & & TCS  \\
Jun 26, 1994 & 49529.5   &  9.22$\pm$0.02 &  9.14$\pm$0.01 &  9.06$\pm$0.03 & & TCS  \\
Jun 28, 1994 & 49531.5   &  9.22$\pm$0.01 &  9.15$\pm$0.01 &  9.03$\pm$0.01 & & TCS  \\
Jun 30, 1994 & 49533.5   &  9.24$\pm$0.01 &  9.09$\pm$0.02 &  9.08$\pm$0.02 & & TCS  \\
Jul 01, 1994 & 49534.5   &  9.17$\pm$0.01 &  9.12$\pm$0.01 &  8.99$\pm$0.00 & & TCS  \\
Jul 02, 1994 & 49535.5   &  9.26$\pm$0.02 &  9.20$\pm$0.01 &  9.09$\pm$0.01 & & TCS  \\
Jul 03, 1994 & 49536.5   &  9.27$\pm$0.02 &  9.17$\pm$0.01 &  9.12$\pm$0.01 & & TCS  \\
Nov 05, 1994 & 49661.5   &  9.16$\pm$0.01 &  9.09$\pm$0.01 &  9.06$\pm$0.01 & & TCS   \\
Nov 06, 1994 & 49662.5   &  9.01$\pm$0.01 &  8.94$\pm$0.01 &  8.90$\pm$0.01 & & TCS  \\
Nov 07, 1994 & 49663.5   &  9.28$\pm$0.02 &  9.21$\pm$0.01 &  9.14$\pm$0.02 & & TCS \\
Nov 08, 1994 & 49664.5   &  9.29$\pm$0.02 &  9.18$\pm$0.01 &  9.08$\pm$0.00 & & TCS \\
Nov 09, 1994 & 49665.5   &  9.21$\pm$0.01 &  9.11$\pm$0.01 &  9.08$\pm$0.01 & & TCS  \\
Jan 02, 1995 & 49719.5   &  9.23$\pm$0.01 &  9.15$\pm$0.01 &  9.03$\pm$0.01 & & TCS  \\
Jan 03, 1995 & 49720.5   &  9.12$\pm$0.01 &  9.07$\pm$0.01 &  9.00$\pm$0.01 & & TCS  \\
Jan 05, 1995 & 49722.5   &  9.14$\pm$0.02 &  9.09$\pm$0.01 &  8.99$\pm$0.01 & & TCS  \\
Apr 24, 1995 & 49831.5   &  9.16$\pm$0.01 &  9.07$\pm$0.01 &  9.02$\pm$0.02 & & TCS  \\
Apr 25, 1995 & 49832.5   &  9.21$\pm$0.01 &  9.12$\pm$0.01 &  9.07$\pm$0.00 & & TCS  \\
Apr 26, 1995 & 49833.5   &  9.25$\pm$0.01 &  9.13$\pm$0.01 &  9.03$\pm$0.01 & & TCS  \\
Apr 28, 1995 & 49835.5   &  9.23$\pm$0.01 &  9.14$\pm$0.01 &  9.05$\pm$0.01 & & TCS  \\
Apr 29, 1995 & 49836.5   &  9.20$\pm$0.01 &  9.08$\pm$0.01 &  9.04$\pm$0.01 & & TCS  \\
Jul 28, 1995 & 49926.5   &  9.22$\pm$0.02 &  9.12$\pm$0.01 &  9.09$\pm$0.15 & & TCS  \\
Jul 28, 1995 & 49926.5   &  9.23$\pm$0.01 &  9.04$\pm$0.01 &  9.60$\pm$0.13 & & TCS \\
Jul 29, 1995 & 49927.5   &  9.18$\pm$0.01 &  9.06$\pm$0.01 &  8.94$\pm$0.06 & & TCS  \\
Jul 29, 1995 & 49927.5   &  9.19$\pm$0.01 &  9.04$\pm$0.01 &  9.09$\pm$0.11 & & TCS  \\
Jul 31, 1995 & 49929.5   &  9.27$\pm$0.01 &  9.10$\pm$0.01 &  8.95$\pm$0.01 & & TCS  \\
Oct 11, 1995 & 50001.5   &  9.36$\pm$0.01 &  9.17$\pm$0.01 &  9.00$\pm$0.00 & & TCS \\
Oct 14, 1995 & 50004.5   &  9.15$\pm$0.01 &  9.03$\pm$0.01 &  8.90$\pm$0.01 & & TCS \\
Jul 27, 1996 & 50291.5   &  9.40$\pm$0.01 &  9.25$\pm$0.01 &  9.07$\pm$0.00 & & TCS  \\
Jul 28, 1996 & 50292.5   &  9.14$\pm$0.01 &  9.03$\pm$0.01 &  8.92$\pm$0.01 & & TCS  \\
Jul 16, 1997 & 50645.5   &  9.20$\pm$0.01 &  9.09$\pm$0.01 &  8.98$\pm$0.01 & & TCS  \\
Jul 17, 1997 & 50646.5   &  9.18$\pm$0.01 &  9.06$\pm$0.01 &  8.96$\pm$0.01 & & TCS  \\
Jul 19, 1997 & 50648.5   &  9.17$\pm$0.01 &  9.07$\pm$0.01 &  8.99$\pm$0.00 & & TCS  \\
Jun 15, 1998 & 50979.5   &  9.16$\pm$0.01 &  9.09$\pm$0.01 &  8.98$\pm$0.01 & & TCS  \\
Jun 16, 1998 & 50980.5   &  9.12$\pm$0.01 &  9.07$\pm$0.01 &  8.98$\pm$0.01 & & TCS \\
Jun 17, 1998 & 50981.5   &  9.19$\pm$0.01 &  9.10$\pm$0.01 &  8.99$\pm$0.00 & & TCS  \\
Oct 27, 1998 & 51113.5   &  9.29$\pm$0.01 &  9.17$\pm$0.01 &  9.13$\pm$0.00 & & TCS  \\
Jul 26, 1999 & 51385.5   &  9.23$\pm$0.02 &  9.12$\pm$0.02 &  9.09$\pm$0.02 & & TCS  \\
Jul 28, 1999 & 51387.5   &  9.20$\pm$0.02 &  9.13$\pm$0.02 &  9.05$\pm$0.02 & & TCS  \\
Jul 31, 1999 & 51390.5   &  9.22$\pm$0.02 &  9.13$\pm$0.02 &  9.06$\pm$0.02 & & TCS  \\
Oct 02, 1999 & 51453.5   &  9.18$\pm$0.02 &  9.08$\pm$0.02 &  8.98$\pm$0.02 & & TCS  \\
Oct 04, 1999 & 51455.5   &  9.20$\pm$0.02 &  9.10$\pm$0.02 &  9.03$\pm$0.02 & & TCS  \\
Oct 05, 1999 & 51456.5   &  9.20$\pm$0.02 &  9.13$\pm$0.01 &  9.02$\pm$0.02 & & TCS  \\
Jan 15, 2001 & 51924.5   &  9.20$\pm$0.02 &  9.08$\pm$0.02 &  8.96$\pm$0.02 & & TCS  \\
\end{longtable}
\end{center}
%------------------------------------------------------------

\end{document}